# Determination of the Dzyaloshinskii-Moriya interaction using pattern recognition and machine learning

Masashi Kawaguchi[1†], Kenji Tanabe[2†*], Keisuke Yamada[3], Takuya Sawa[2], Shun Hasegawa[1], Masamitsu Hayashi[1**], and Yoshinobu Nakatani[4***]

*[1]Department of Physics, The University of Tokyo, Tokyo, 113-0033, Japan*
*[2]Toyota Technological Institute, Nagoya, 468-8511, Japan*
*[3]Department of Chemistry and Biomolecular Science, Faculty of Engineering, Gifu University, Gifu, 501-1193, Japan*
*[4]University of Electro-Communications, Chofu, 182-8585, Japan*

*[†]Equally contributed*
*[*]tanabe@toyota-ti.ac.jp, [**]hayashi@phys.s.u-tokyo.ac.jp, [***]nakatani@cs.uec.ac.j*

**ABSTRACT**

Machine learning is applied to a large number of modern devices that are essential in building energy efficient smart society. Audio and face recognition are among the most well-known technologies that make use of such artificial intelligence. In materials research, machine learning is adapted to predict materials with certain functionalities, an approach often referred to as materials informatics. Here we show that machine learning can be used to extract material parameters from a single image obtained in experiments. The Dzyaloshinskii-Moriya (DM) interaction and the magnetic anisotropy distribution of thin film heterostructures, parameters that are critical in developing next generation storage class magnetic memory technologies, are estimated from a magnetic domain image. Micromagnetic simulation is used to generate thousands of random images for training and model validation. A convolutional neural network system is employed as the learning tool. The DM exchange constant of typical Co-based thin film heterostructures is studied using the trained system: the estimated values are in good agreement with experiments. Moreover, we show that the system can independently determine the magnetic anisotropy distribution, demonstrating the potential of pattern recognition. This approach can considerably simplify experimental processes and broaden the scope of materials research.

## 1. INTRODUCTION

The Dzyaloshinskii-Moriya (DM) interaction[1,2] is an antisymmetric exchange interaction that favors non-collinear alignment of magnetic moments and induces chiral magnetic order. In contrast to the Heisenberg exchange interaction that forms the basis of ferromagnetic and antiferromagnetic orders, the DM interaction is the source of unconventional magnetic textures. For example, spin spirals[3], chiral Néel domain walls[4,5], and skyrmions[6-8] have been observed in bulk and thin film magnets with strong DM interaction. Importantly, chiral Néel domain walls and skyrmions can be driven by spin current that diffuses into the magnetic layer via the spin Hall effect of neighboring non-magnetic layers[9-13]. Such magnetic objects are topologically protected from annihilating each other[14,15], a property that is absent in other magnetic systems. Current controlled motion of chiral Néel domain walls[10,11] and skyrmions[12,13] are thus attracting significant interest for their potential use in storage class magnetic memories[16-19].

Recent studies have shown that the DM interaction emerge at the interface of ferromagnetic layer and non-magnetic layer with strong spin orbit interaction[17]. Although the underlying mechanism of such interfacial DM interaction is under debate, its size is sufficiently large to stabilize chiral domain walls and isolated skyrmions. To evaluate the strength and chirality of the DM interaction, i.e. the DM exchange constant, a number of approaches have been proposed. As many of the approaches make use of the dynamics of the magnetic system, for example, current or field induced motion of domain walls[10,11,20,21], propagation of spin waves[22] and current/field dependence of the magnetization reversal processes[23,24], there are difficulties in accurately extracting the DM exchange constant. The difficulties arise in part because the value depends on the model used to describe the system. In addition, random pinning of domain walls (and spin textures), which originates from the magnetic anisotropy distribution within the magnetic thin film, influences magnetization dynamics and adds uncertainty in the determination of the DM exchange constant. Since there is almost no means to control (and evaluate) the magnetic anisotropy distribution, estimation of DM exchange constant relies on the given property of each system.

Magnetic domain structure at equilibrium is determined by minimization of magnetic energy of the system, which typically includes magneto-static, magneto-elastic, anisotropy, Heisenberg exchange and DM exchange energies. The pattern of the magnetic domain structure therefore includes information of the DM exchange constant. Recent studies have shown that the radius of skyrmions allow determination of its size[13,25-27]. As the size of the skyrmions is of the order of few tens of nanometers, however, it remains as a significant challenge to obtain their images with typical laboratory equipment. Similarly, mapping the magnetization direction of magnetic domain walls, which are typically a few nanometer wide, requires state of the art imaging techniques[5,28].

Here we show that the DM exchange constant can be simply extracted from a micrometer-scale magnetic domain image using pattern recognition and machine learning. A convolutional neural network is used to characterize the magnetic domain pattern. To train the neural network, a large number of images with different patterns that derive from a magnetic system with fixed material parameters are required. As it is extremely challenging to synthesize films with well-defined material parameters, here we use micromagnetic simulations to generate the images for supervised learning. Micromagnetic simulation is a widely used tool to study magnetic systems. The simulation is capable of returning images that resemble those obtained in the experiments[13,29,30]. The system is trained and tested using the images generated from the simulations. As a demonstration, we use the trained system to estimate the DM exchange constant from experimentally obtained magnetic domain images (see Fig. 1 for the procedure used). We show that the trained system can estimate not only the DM exchange constant, which is in good agreement with experiments, but also the distribution of the magnetic anisotropy energy, for which only a few experimental studies have been reported thus far[31,32].

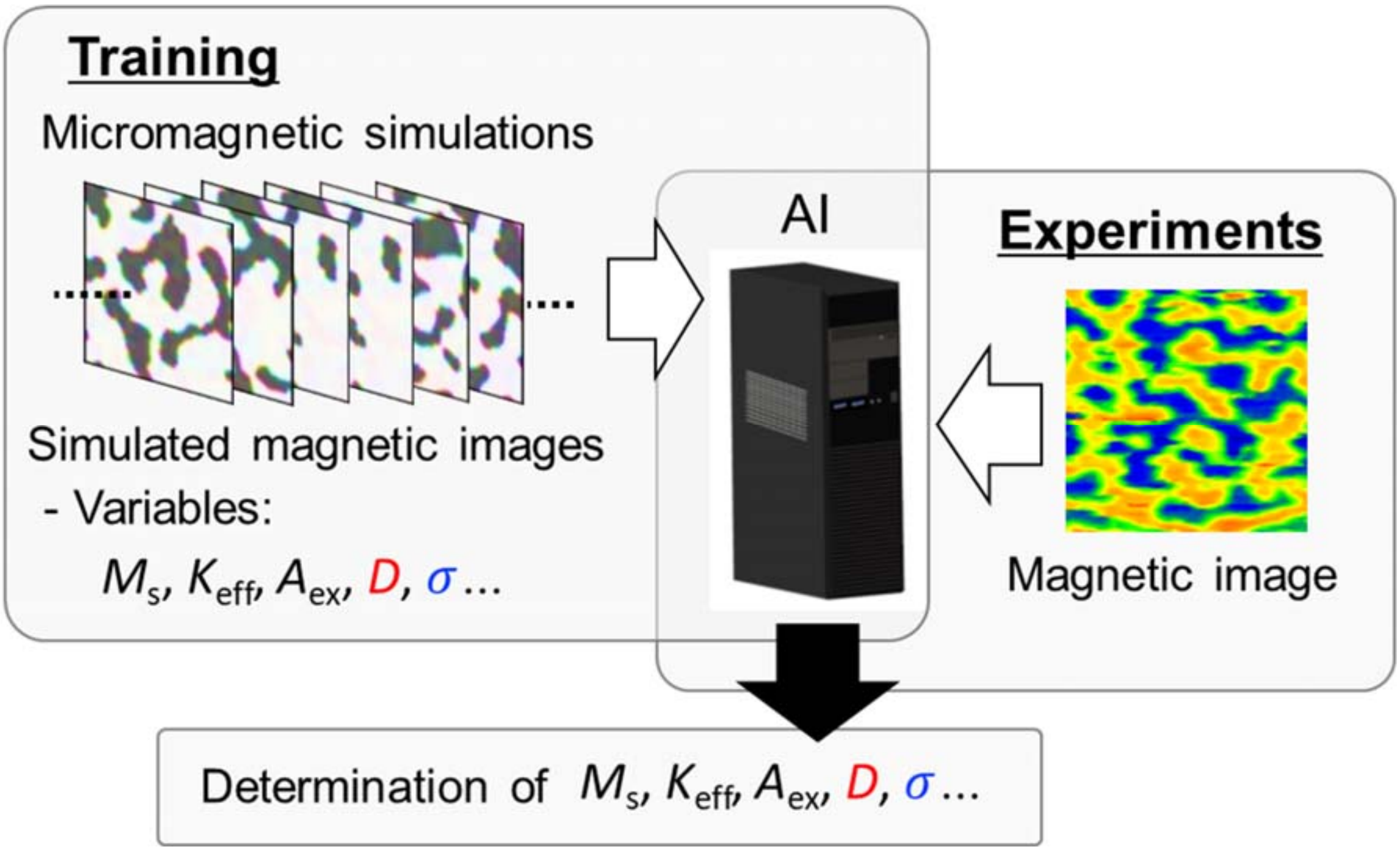

**Fig. 1. Concept of the study.** Micromagnetic simulations are used to generate thousands of training images. There are five relevant material parameters: $M_s$, $K_{eff}$, $A_{ex}$, $D$ and $\sigma$. Here we vary $D$ and $\sigma$ in the simulations so that the system can learn domain patterns with different $D$ and $\sigma$. After the supervised training, we feed the system with experimentally obtained image of magnetic domains to extract $D$ and $\sigma$.

## 2. RESULTS AND DISCUSSIONS

### 2.1 Preparation of training and testing data sets

The training data set is generated using a homemade micromagnetic simulation code. See Methods for the details of the calculations. The magnetic anisotropy dispersion ($\sigma$) is defined as $\sigma \equiv \Delta K_\mathrm{u}/K_\mathrm{u}$, where $K_\mathrm{u}$ is the magnetic anisotropy energy density and $\Delta K_\mathrm{u}$ is its variation. We first generate training images with fixed $\sigma$ ($\sigma = 0.15$), $K_\mathrm{u}$, saturation magnetization ($M_s$) and exchange constant ($A_\mathrm{ex}$). The value of each parameter is chosen to mimic typical thin film heterostructures[33] (see Table 1). The DM exchange constant ($D$) is varied from 0 to 1.00 mJ $m^{-2}$. The initial condition and the pattern of $K_\mathrm{u}$ distribution are varied to generate 100,000 training images of the equilibrium magnetic state for a given parameter set with various values of $D$ (see Methods and the Supplementary Fig. 1 for the details). Exemplary images of the equilibrium magnetic state with different $D$ are shown in Fig. 2(a). The domain size tends to shrink with increasing $D$, consistent with theoretical models[13,25-27]. Due to a non-zero $\sigma$ which causes random pinning, it is difficult to identify a clear trend in the shape of the domains with varying $D$.

Model validation is performed with 10,000 testing images with different values of $D$ created using the same code. $D^\mathrm{set}$ corresponds to $D$ used in the simulations to generate the testing images. The testing images are studied by the trained system: the estimated $D$ returned from the system is denoted as $D^\mathrm{est}$. The relation of $D^\mathrm{set}$ vs. $D^\mathrm{est}$ is shown in Fig. 2(b). When $D^\mathrm{set}$ is larger than ~0.05 mJ $m^{-2}$, we find a linear relation between $D^\mathrm{set}$ vs. $D^\mathrm{est}$ with a root mean square (rms) error of ~0.046 mJ $m^{-2}$. To show the distribution of $D^\mathrm{est}$ more clearly, we plot the histogram of $D^\mathrm{est}$ for 5 different values of $D^\mathrm{set}$ (Fig. 2 (c)). The standard deviation of each histogram is ~0.05 mJ $m^{-2}$, consistent with the rms error of $D^\mathrm{set}$ vs. $D^\mathrm{est}$. These results show that the system cannot accurately determine $D$ when $D^\mathrm{set} \lesssim 0.05$ mJ $m^{-2}$.

In experiments, it is typically the case that $\sigma$ is not a known parameter. It is therefore more effective if one can determine both parameters, $D$ and $\sigma$, at once from a single magnetic domain image. We have thus created training images with both $D$ and $\sigma$ varied. On top of the changes in $D$ (0 ~ 1.00 mJ $m^{-2}$), we vary $\sigma$ from 0 to 0.2. We generate 100,000 testing images with various values of $D$ and $\sigma$, different initial condition and anisotropy distribution pattern. Figure 2(d) shows images of the equilibrium magnetic states with different $D$ and $\sigma$. As the value of $\sigma$ is random here, we find almost no trend in the size as well as the shape of the domains with increasing $D$. Although human eyes can hardly identify any pattern associated with changes in $D$, the trained system does a surprising good job in detecting the difference. Again, we generate 10,000 testing images using the same code for model validation. In Fig. 2(e), we show the $D^\mathrm{set}$ dependence of $D^\mathrm{est}$. Albeit the variation in $\sigma$, $D^\mathrm{est}$ scales $D^\mathrm{set}$ with a rms error of ~0.045 mJ $m^{-2}$, nearly the same with that of the training data set with a fixed $\sigma$ (Fig. 2(b)). These results show that the trained system does not rely on the size of magnetic domains to determine $D$: we infer that the curvature of the domains as well as the shape of the domain boundary play a role in the determination process[25].

Interestingly, the trained system can independently determine the value of $\sigma$ in addition to $D$. The estimated value of $\sigma$ ($\sigma^\mathrm{est}$) is plotted against the set value $\sigma^\mathrm{set}$ in Fig. 2(f). As evident, the trained system provides accurate estimation of $\sigma$: the rms error is ~0.005. These results clearly show that trained system can estimate multiple material parameters simultaneously from a single magnetic domain image. Provided that the parameters are not correlated, we consider the approach can be extended to estimate other parameters (e.g. $K_\mathrm{eff}$, $M_s$, $A_\mathrm{ex}$) as well.

**Table 1.** Summary of the material parameters used in micromagnetic simulations.

| | $t$ (nm) | $M_\mathrm{s}$ (kA $m^{-1}$) | $K_\mathrm{eff}$ ($10^5$ J $m^{-3}$) | $A_\mathrm{ex}$ ($10^{-11}$ J $m^{-1}$) | $D$ (mJ $m^{-2}$) | $\sigma$ | $T$ (K) |
|---|---|---|---|---|---|---|---|
| Fig. 2(a-c) | 1.2 | 1500 | 1.9 | 3.1 | 0~1.0 | 0.15 | 0 |
| Fig. 2(d-f) | 1.2 | 1500 | 1.9 | 3.1 | 0~1.0 | 0.00~0.2 | 0 |
| Fig. 4 | 0.9 | 1445 | 2.4/3.7 | 4 | 0~1.5 | 0.05~0.2 | 300 |

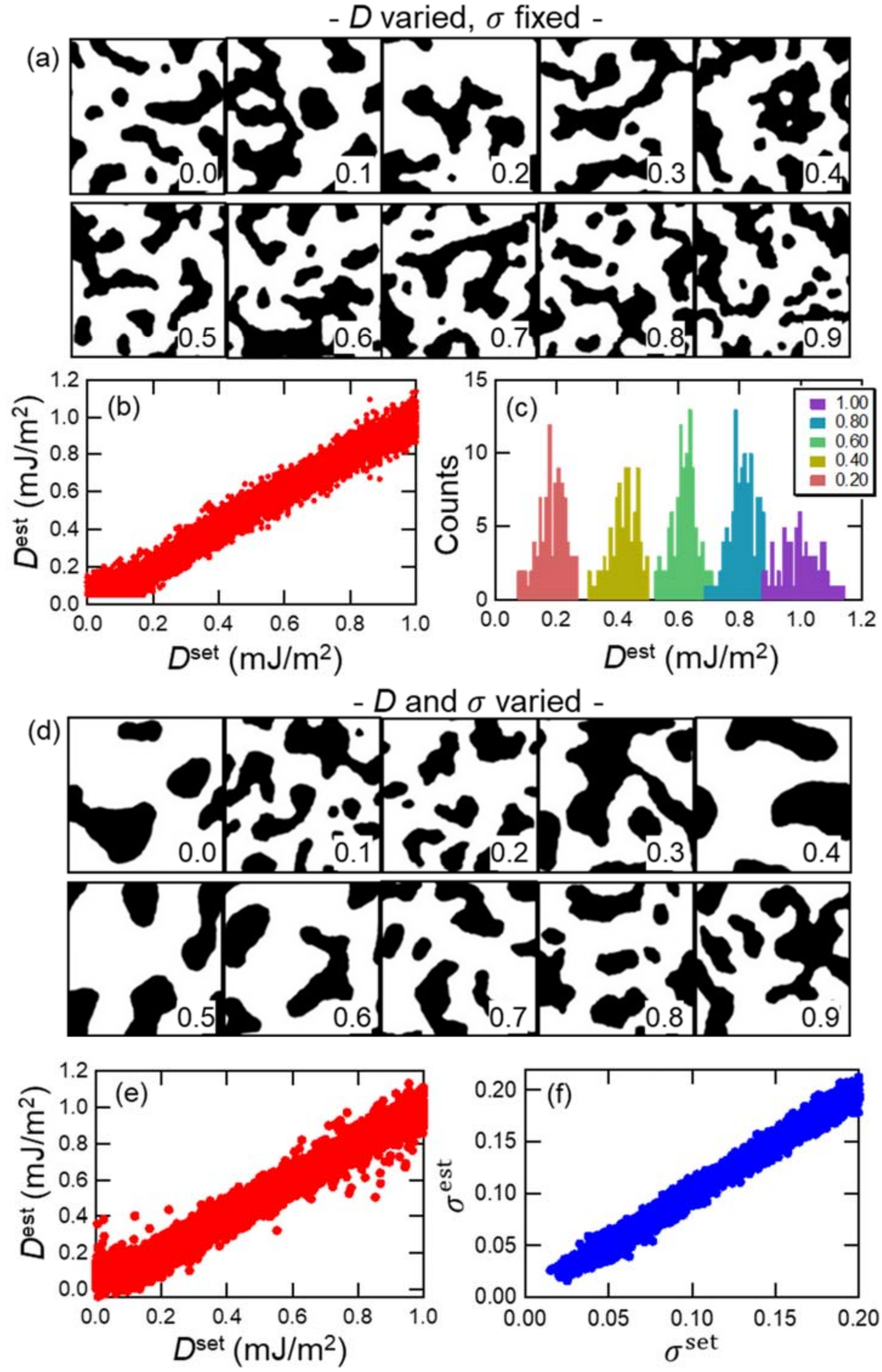

**Fig. 2. Training and validation of the neural network system.** (a) Typical magnetic domain images calculated using micromagnetic simulations (the images are used for training). Dark and bright contrast represents the magnetization direction along the film normal. The DM exchange constant is varied from 0 to 0.90 mJ $m^{-2}$: the corresponding value is indicated at the bottom right corner of each image. (b) The DM exchange constant ($D^{\mathrm{est}}$) estimated from the testing images are plotted as a function of $D$ set in the simulations ($D^{\mathrm{set}}$). (c) Histograms of the $D^{\mathrm{est}}$ for $D^{\mathrm{set}} = 0.2, 0.4, 0.6, 0.8, 1.0$ mJ $m^{-2}$. (a-c) $\sigma$ is fixed to 0.15 in the simulations to generate the training/testing images. (d) Same with (a) except that, in addition to $D$, $\sigma$ is randomly varied in the process of creating training images. The images shown are randomly chosen from a set of training images with fixed $D$ but various $\sigma$. The corresponding value of $D$ is indicated at the bottom right corner of each image. (e) $D^{\mathrm{est}}$ vs. $D^{\mathrm{set}}$. (f) $\sigma^{\mathrm{est}}$ vs. $\sigma^{\mathrm{set}}$. (d-f) Both $D$ and $\sigma$ are varied in the simulations to generate the training/testing images. See Table 1 for the values of all material parameters.

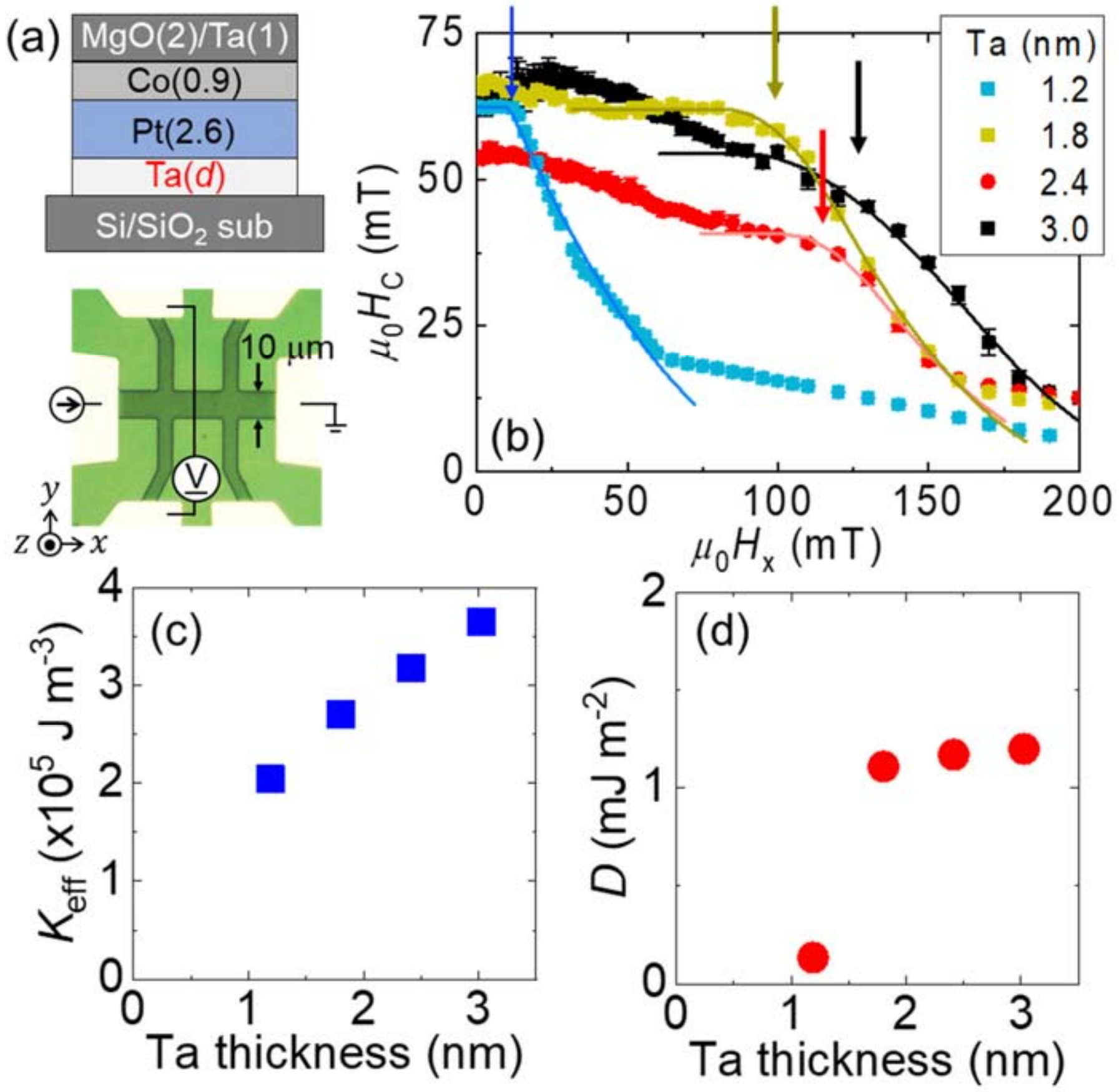

**Fig. 3. Experimental determination of the DM interaction.** (a) Schematic illustration of the film structure, an optical microscope image of a representative device and definition of the coordinate axes. (b) Switching field $H_c$ plotted as a function of $H_x$ for Hall bars made from heterostructures with different Ta seed layer thickness (*d*). The vertical arrows indicate $H_{DM}$ obtained by fitting the data with model calculations[24]. The solid lines show the fitting results. (c,d) *d* dependence of the effective perpendicular magnetic anisotropy energy ($K_{eff}$) (c) and the DM exchange constant (*D*) (d).

### 2.2 Magnetic properties of the samples for pattern recognition.

We next use the trained system to estimate $D$ and $\sigma$ from experimentally obtained magnetic domain images. The film of the samples used is: Si sub./Ta (*d*)/Pt (2.6 nm)/Co (0.9 nm)/MgO (2 nm)/Ta (1 nm). Details of sample preparation and characterization are described in Methods. The thickness of the Ta seed layer ($d$) is varied to change $D$ of the films via modification of the (111) texture of the Pt layer (while attempting to minimize changes to other parameters). The magnetic easy axis of the films points along the film normal. The average $M_s$ of the measured films is ~1445 kA m$^{-1}$. The $d$ dependence of $K_{eff}$, i.e. the effective magnetic anisotropy energy density defined as $K_{eff} = K_u - \frac{M_s^2}{2\mu_0}$ ($\mu_0$ is the vacuum permeability), is shown in Fig. 3(c). The increase in $K_{eff}$ with increasing $d$ is associated with the improvement of the texture of the Pt and Co layers.

The DM exchange constant is estimated using magnetic field induced switching of magnetization[24]. A Hall bar is patterned from the films using conventional optical lithography. See Fig. 3(a) for a schematic illustration of the film structure, an optical microscope image of a representative device and definition of the coordinate axis. We use the Hall voltage to probe the *z*-component (i.e. along the film normal) of the magnetization via the anomalous Hall effect. To extract $D$, the easy axis switching field ($H_C$) is studied as a function of in-plane magnetic field ($H_x$). Fig. 3(b) shows the $H_x$ dependence of $H_C$ for films with different $d$. As reported previously[24], in systems with non-zero $D$, $H_C$ shows a sharp decrease with increasing $|H_x|$ at $H_x{\sim}H_{DM}$, where $H_{DM}$ is the DM exchange field defined as $H_{DM} = D/(\mu_0 M_s \Delta)$ ($\Delta = \sqrt{A_{ex}/K_{eff}}$). The data is fitted with a model calculation[24] to obtain $H_{DM}$: the results are shown by the solid lines in Fig. 3(b). The $d$ dependence of $D$ is plotted in Fig. 3(d). We find $D$ shows a sharp increase as $d$ exceeds ~1 nm. We consider the texture of the Pt/Co interface plays a dominant role in defining $D$. The size of $D$ when $d$ exceeds ~1 nm is in agreement with past reports[20,34-36] (see Supplementary Table 1 for values of $D$ obtained in similar heterostructures).

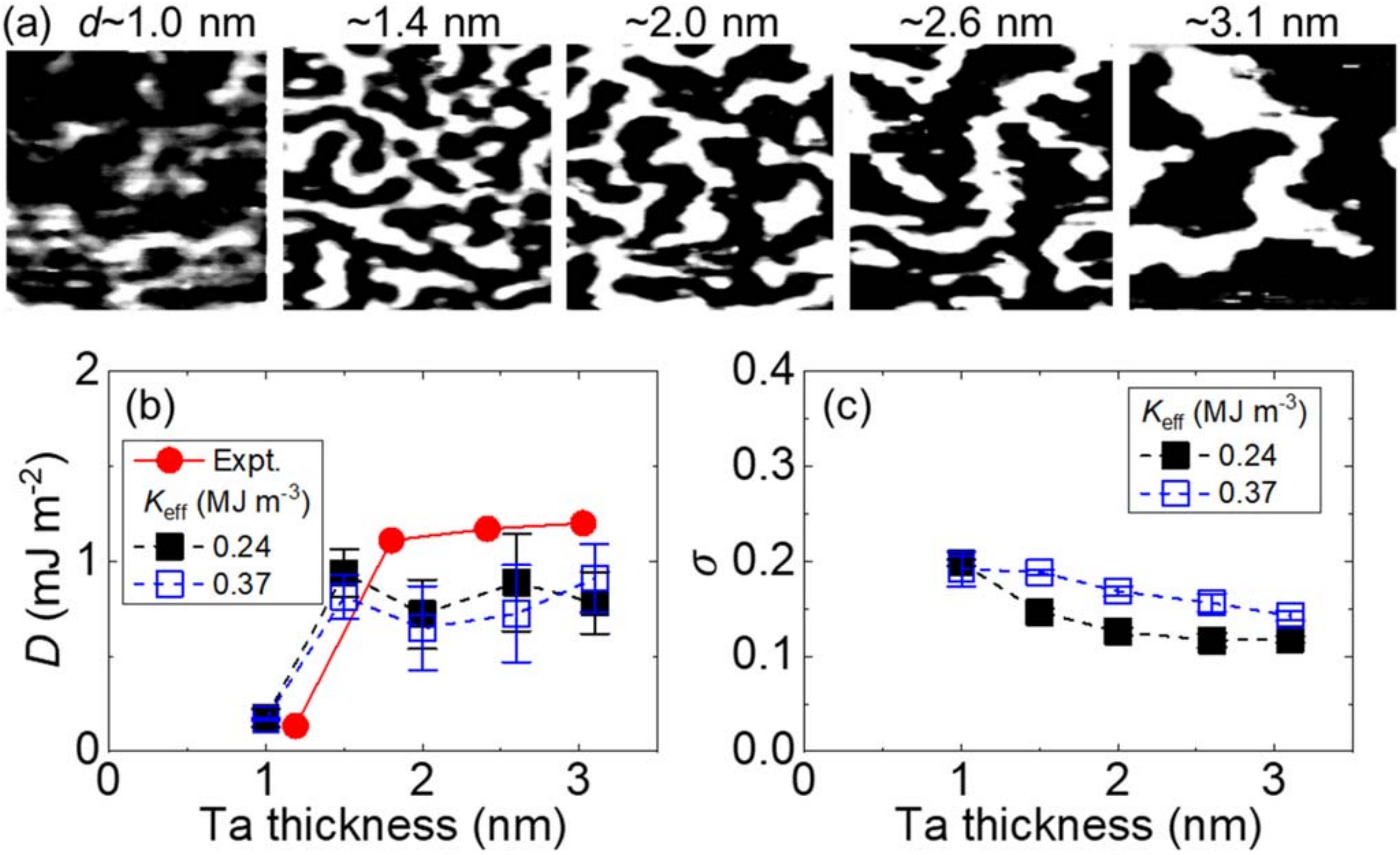

**Fig. 4. $D$ and $\sigma$ obtained from pattern recognition and machine learning.** (a) Experimentally obtained magnetic domain images using a magnetic microscope equipped with a MTJ sensor. The bright and dark contrast represent magnetization pointing from and into the paper. The thickness of the Ta seed layer ($d$) is denoted in each image. (b,c) DM exchange constant ($D$) (b) and distribution of $K_u$ ($\sigma$) (c) estimated from the domain images using the trained system. Two different values of $K_{eff}$ are used in the simulations to generate the training images: the estimated values ($D$ and $\sigma$) obtained from the trained systems are denoted using open and solid squares. The error bars show 95% confidence interval (see Methods). $D$ from Fig. 3(c) are shown together by the red circles in (b).

### 2.3 Pattern recognition of magnetic domain images.

The magnetic domain images of the films are acquired using a magnetic microscope equipped with a magnetic tunnel junction (MTJ) sensor[37]. We note that a more common Kerr microscopy can be used for the imaging. Here we are limited by the size of the training images generated by micromagnetic simulations: to save computation time, we have used images with dimension of ~2 × 2 μm$^2$. Since the neural network is trained using these images, the spatial resolution of the imaging tool must be significantly better than ~1 μm, which excludes the use of conventional Kerr microscopy.

Typical magnetic domain images obtained using the microcopy are shown in Fig. 4(a). Clearly, the size of the domains changes as $d$ is varied. These images are fed into the trained system to estimate $D$ and $\sigma$. To mimic the experimental condition, $M_s$ and $K_{eff}$ used in the simulations to generate training images are chosen from experiments and $A_{ex}$ is taken from past reports on similar systems[38,39]. Since $K_{eff}$ varies with $d$ (see Fig. 3(c)), we use the lower (~0.24 MJ m$^{-3}$) and upper (~0.37 MJ m$^{-3}$) limits of $K_{eff}$ in the simulations. We have also included Langevin field in the simulation to emulate thermal fluctuation (see Methods). The value of $D$ the trained system returned for each image is plotted against $d$ in Fig. 4(b). Interestingly the $d$ dependence of the estimated $D$ is consistent with that of the experiments (red circles in Fig. 4(b)). Note that the magnitude of $K_{eff}$ does not significantly influence estimation of $D$. Values of $\sigma$ obtained from the images are plotted as a function of $d$ in Fig. 4(c). The size of $\sigma$ estimated from the images can be compared to that obtained, for example, via measurements of the domain wall velocity distribution ($\sigma$~0.15)[31] and magnetic hysteresis loops of a nano-patterned structure[32]. We find $\sigma$ tends to monotonically decrease with increasing $d$. This is in sharp contrast to $D$, which shows an abrupt increase when $d$ exceeds ~1 nm. The monotonic change of $\sigma$ with $d$ is in accordance with that of $K_{eff}$ (Fig. 3(c)). Similar to $K_{eff}$, we infer that $\sigma$ is related to the texture of the Pt/Co layer, however, in a different way than that of $D$. The stark difference in the $d$ dependence of the estimated $D$ and $\sigma$ demonstrates that the trained system can identify multiple parameters independently as long as they are not correlated.

## 3. CONCLUSION

In summary, we have demonstrated that pattern recognition and machine learning can be applied to extract critical material parameters from a single magnetic domain image. In particular, we show that the DM exchange constant ($D$) and distribution of the magnetic anisotropy energy ($\sigma$), two key parameters that are difficult to assess experimentally, can be extracted from an image. The accuracy of the supervised learning in estimating $D$ and $\sigma$ are found to be ~0.05 mJ $m^{-2}$ and ~0.005, respectively, which can likely be reduced with improved learning algorithms. As a proof of concept, we use the trained system to estimate $D$ and $\sigma$ of Co-based heterostructures using magnetic domain images obtained from a magnetic microscope. The estimated value of $D$ is in good agreement with that extracted from experiments. This approach can be extended to estimate all relevant material parameters (e.g. $M_s$, $K_{\mathrm{eff}}$, $A_{\mathrm{ex}}$,…) at once from a single magnetic domain image, which will significantly simplify materials research for magnetic memory and storage technologies.

## METHODS

### Sample preparation and film characterization

Films are deposited using rf magnetron sputtering on silicon substrate. The film structure is: Si sub./Ta ($d$)/Pt (2.6 nm)/Co (0.9 nm)/MgO (2 nm)/Ta (1 nm). A moving shutter is used to vary the thickness ($d$) of the Ta seed layer linearly across the substrate. $d$ is varied from ~0 to ~3 nm across a 10 mm long substrate. The MgO (2 nm)/Ta (1 nm) serves as a capping layer to prevent oxidation of the Co layer. The saturation magnetization ($M_s$) of the heterostructure is studied using vibrating sample magnetometry (VSM). We take the average value of $M_s$ for films with varying $d$.

The heterostructure is patterned into Hall bars using optical lithography and Ar ion etching. The length and width of the current channel of the Hall bar is ~60 μm and ~10 μm, respectively. Contact pads made of Ta (5 nm)/Cu (60 nm)/Pt (20 nm) are formed using optical lithography and liftoff. The effective magnetic anisotropy field ($H_{\mathrm{K}}$) is obtained via transport measurements. The Hall resistance is measured under application of in-plane magnetic field. The field at which the Hall resistance saturates is defined as $H_{\mathrm{K}}$. Except for films with $d$ less ~0.5 nm, we find the magnetic easy axis of the heterostructure points along the film normal.

### Machine learning

The convolutional neural network (CNN) system used in this paper contains twelve layers. The first ten are convolution and the remaining two are fully connected. The filter size of the convolution layers is 3×3, and the strides of the first six and the last four layers are 1×1 and 2×2, respectively. The number of the filters for each convolution layer is 64, 64, 40, 36, 32, 28, 24, 20, 16 and 16, respectively. The number of the units of the first fully connected layer is 10. The ReLU is applied to the output of all convolutions and the first fully-connected layer. The Huber loss is used for loss calculation. The Adam algorithm is used for optimization. The network was trained using a commercial deep learning tool, Sony Neural Network Console, with batch size of 64 for 100 epochs (https://dl.sony.com/app/).

The number of the testing images is fixed to 1/10 of the training images. For a given training data set, four machines are developed (since the order of the learning process is randomized, the results can be different even though the training data set used is the same). The material parameters estimated from the four machines are used to obtain the mean value and standard deviation. See Supplementary Figs. 2 and 3 for the details as well as the effect of cross validation and data augmentation on the machine learning performance. To estimate $D$ from the experimental images (Fig. 4), we augment data with image rotation. Each image is rotated 90 deg, 180 deg and 270 deg to generate three additional images. The four images are fed into the four machines to obtain 16 values of $D$. The average value of the 16 data are shown in Fig. 4. The 95% confidence interval is calculated using the mean and the variance of the 16 data.

In addition to the convolutional neural network (CNN) system, we have tested a simple residual network (RN) network. See Supplementary Fig. 3 for the performance of the simple RN network.

### Micromagnetic simulations

All micromagnetic simulations were performed using a GPU based program developed previously[40,41]. The sample was divided into identical rectangular cells in which magnetization was assumed to be constant. The motion of magnetization was calculated by solving the Landau-Lifshitz-Gilbert equation with thermal noise[42] (i.e. the Langevin equation).

$$\frac{\partial \hat{m}}{\partial t} = -\gamma \hat{m} \times \vec{H}_{\mathrm{eff}} + \alpha \hat{m} \times \frac{\partial \hat{m}}{\partial t}$$

Here, $\gamma$, $\hat{m}$, $\vec{H}_{\mathrm{eff}}$ and $\alpha$, are the gyromagnetic ratio, a unit vector representing the magnetization direction, the effective magnetic field and the Gilbert damping constant.

$\vec{H}_{\text{eff}}$ is calculated from the magnetic energy $\varepsilon$:

$$\vec{H}_{\text{eff}} = -\frac{\delta\varepsilon}{\delta\vec{M}} + \vec{h}(t)$$

$$\varepsilon = \varepsilon^A + \varepsilon^K + \varepsilon^{DM} - \frac{1}{2}M_s\hat{m}\cdot\vec{H}^D$$

$$\varepsilon^A = A_{\text{ex}}(\nabla\hat{m})^2$$

$$\varepsilon^K = K_{\text{u}}(1 - m_z^2)$$

$$\varepsilon^{DM} = D\left[\left(m_x\frac{\partial m_z}{\partial x} - m_z\frac{\partial m_x}{\partial x}\right) + \left(m_y\frac{\partial m_z}{\partial y} - m_z\frac{\partial m_y}{\partial y}\right)\right]$$

$$< h_i(t)h_j(t+\tau) >= \frac{2kT\alpha}{|\gamma|vM_s}\delta(\tau)\delta_{ij}$$

Here $\vec{h}(t)$ is the effective field associated with thermal energy, $\varepsilon^A$, $\varepsilon^K$ and $\varepsilon^{DM}$ are the exchange energy, the anisotropy energy and the Dzyaloshinskii-Moriya (DM) interaction energy, and $\vec{H}^D$ is the demagnetizing field. $M_s$, $A_{\text{ex}}$, $K_u$, $D$, $k$, $T$, $v$, $\delta(\tau)$, and $\delta_{ij}$ are the saturation magnetization, the exchange stiffness constant, the uniaxial magnetic anisotropy energy density, the DM exchange constant, the Boltzmann constant, temperature, the volume of the cell, the Dirac delta function, and the Kronecker delta, respectively. Note that $K_{\text{eff}} = K_{\text{u}} - M_s^2/(2\mu_0)$. The demagnetizing field is calculated numerically.

We vary the pattern of anisotropy distribution and the initial magnetization configuration to generate images with different domain structures. These patterns (anisotropy distribution and the initial magnetization configuration) are created using random number generators. For the former (anisotropy distribution), square groups of 8 × 8 cells were assigned a local anisotropy value randomly distributed according to a normal law with mean $K_{\text{u}}$ and standard deviation σ. For the latter (initial magnetization configuration), a random domain structure is created by uniform random numbers (-1 to 1) for each magnetization components ($m_x$, $m_y$, $m_z$) of each cell. Energy minimization is used to find the equilibrium state at $T$ = 0 K (Fig. 2) or at $T$ = 300 K (Fig. 4). $\alpha$ is set to 1.0 to minimize computation time. Exemplary simulated images of the equilibrium state, with a fixed material parameter set, are shown in Supplementary Fig. 1. The materials parameters used in the calculations are summarized in Table 1.

The dimension of the cell is $4 \times 4 \times t$ nm$^3$. $t$ is the thickness of the ferromagnetic layer. The number of the cell is $512 \times 512$ (the image size is $2.048 \times 2.048$ μm$^2$). Periodic boundary condition is employed to avoid effects from the edges. 110,000 simulations were carried out to produce the results presented in Fig.2. To save computation time for machine learning, each image is converted to a $128 \times 128$ pixels image (one pixel size is $16 \times 16$ nm$^2$) using an averaging filter. Of the 110,000 images generated, 100,000 are used for training and the remaining 10,000 are used for validation (testing). For the results presented in Fig.4, 27,500 simulations were performed and each image is rotated 90°, 180° and 270° to obtain 110,000 images (see Supplementary Fig. 4 for the effect of data augmentation). Again, each image is converted to 128 × 128 pixels to match the pixel size of the image obtained in the experiments.

## DATA AVAILABILITY

The data that support the findings of this study are available from the corresponding author upon reasonable request.

## CODE AVAILABILITY

The code used for the machine learning can be found in https://dl.sony.com/app/.

## ACKNOWLEDGEMENTS

The authors thank H. Awano and S. Sumi for technical support. This work was partly supported by JSPS Grant-in-Aid (grant number: JP19H02553), the Center of Spintronics Research Network of Japan.

## AUTHOR CONTRIBUTIONS

M. K., K.T. and M. H. conceived the experiments. M.K. and M. H. made the samples and measured the material parameters. K. T., K. Y. and T. S. obtained magnetic domain images. Y. N. made the micromagnetic simulation program and performed the micromagnetic simulations. Y. N. and S. H. performed the machine learning. M. K., K. T.,

K. Y., M. H., and Y. N. wrote the main manuscript text and made figures. All authors reviewed the manuscript.

**COMPETING INTERESTS**
The authors declare no competing interests.

**ADDITIONAL INFORMATION**
**Supplementary information** is shown on the following.

**Correspondence** and requests for materials should be addressed to K. T., M. H., and Y. N.

# SUPPLEMENTARY INFORMATION

**Supplementary Note 1:**

**Images from micromagnetic simulations**

Exemplary images from micromagnetic simulations are presented in Supplementary Figure 1 to illustrate the difference in the magnetic domain patterns generated under same material parameters but with different initial condition and different patterns of the magnetic anisotropy distribution (temperature ( $T$ ) is set to 300 K). In Supplementary Figure 1(a), five images generated with $D_{\mathrm{set}} = 0.5$ mJ m$^{-2}$ and $\sigma = 0$ are shown. Here the initial condition (random pattern of magnetization) is varied. Energy minimization is used to find the equilibrium state at $T = 300$ K. Supplementary Figure 1(b) shows five images with $D_{\mathrm{set}} = 0.5$ mJ m$^{-2}$ and $\sigma = 0.15$. Since $\sigma$ is non-zero, in addition to the initial condition, patterns of the magnetic anisotropy distribution are varied. Clearly, the magnetic domain size and the domain patterns are different for Supplementary Figures 1(a) and 1(b).

The estimated DM exchange constant using machine learning are denoted at the top of each image. Although the domain patterns are significantly different, the machine learning returns values that are in good agreement with the set value.

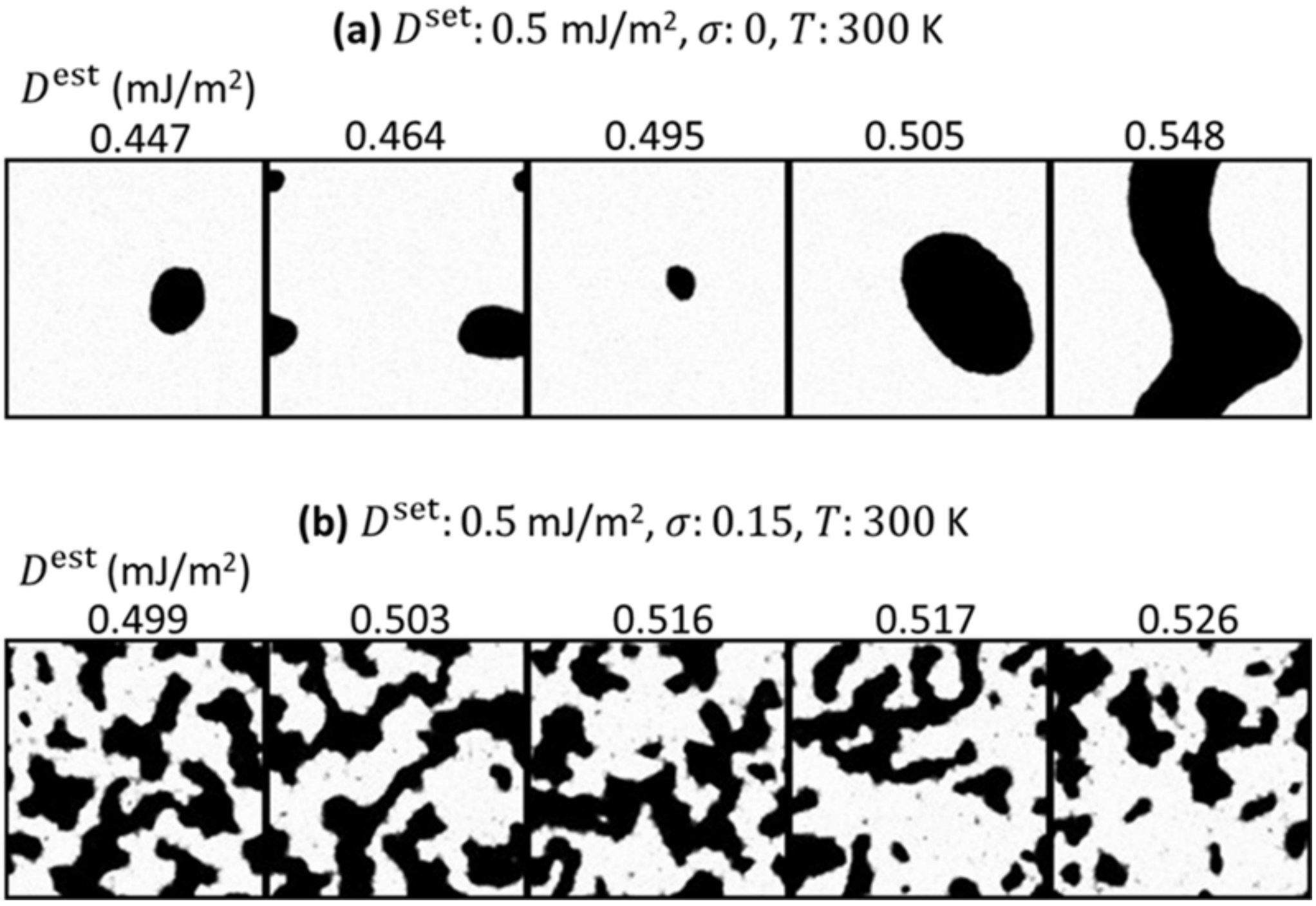

**Supplementary Figure 1: Exemplary images from micromagnetic simulations.** (a,b) Equilibrium magnetic domain images obtained from micromagnetic simulations using $D^{\mathrm{set}} = 0.5$ mJ m$^{-2}$, $\sigma^{\mathrm{set}} = 0$ (a) and $D^{\mathrm{set}} = 0.5$ mJ m$^{-2}$, $\sigma^{\mathrm{set}} = 0.15$ (b). The system temperature is set to $T = 300$ K and the other material parameters are described in the methods section of the main text (Fig. 4). The initial magnetization configuration (a) and the patterns of the magnetic anisotropy distribution are varied (b). The estimated DM exchange constant ($D^{\mathrm{est}}$) using machine learning and pattern recognition are shown at the top of each image. We use image rotation to augment data (4 images from one original) and feed the images to four machines: we take the average of the 16 $D^{\mathrm{est}}$ values.

**Supplementary Note 2:**

**Summary of reported DM exchange constant**

The DM exchange constant ($D$) for Pt/Co/oxide heterostructures reported in the literature are summarized in Supplementary Table 1.

**Supplementary Table 1.** DM exchange constant ($D$) of Pt/Co($t$)/oxide based heterostructures estimated using different methods [1]. Since $D$ depends on the thickness ($t$) of the Co layer (units in nanometer), $D_S \equiv D\ t$ is shown together. BLS stands for Brillouin light scattering. BLS exploits asymmetry in spin wave propagation to study the DM exchange constant.

| Material system (nm) | Method | $D$ (mJ m$^{-2}$) | $D_S$ (pJ m$^{-1}$) | Ref. |
|---|---|---|---|---|
| Pt/Co(0.9)/MgO | Field driven motion of DWs | 1.2 | 1.1 | This work |
| Pt/Co(0.8)/AlOx | Field driven motion of DWs | 1.6~1.9 | 1.3~1.5 | [2] |
| Pt/Co(0.8)/AlOx | Field driven motion of DWs | 0.33 | 0.27 | [3] |
| Pt/Co(0.6)/AlOx | Field driven domain nucleation | 2.2 | 1.3 | [4] |
| Pt/Co(0.5)/MgO | Field driven domain nucleation | 0.48 | 0.24 | [5] |
| Pt/Co(0.93)/AlOx | Current induced motion of DWs | 0.54 | 0.5 | [6] |
| Pt/Co(1)/MgO | Current induced magnetization switching | 2.64 | 2.64 | [7] |
| Pt/Co(0.6-1.2)/AlOx | BLS | 1.6-2.7 | 1.6-1.9 | [8] |
| Pt/Co(1.06)/MgO | BLS | 2.05 | 2.17 | [9] |
| Pt/Co(1)/MgO | BLS | 1.59 | 1.59 | [10] |
| Pt/Co(0-2)/AlOx | BLS | 0.8-1.2 | 1.2-1.8 | [11] |

## Supplementary Note 3:

### Characteristics of machine learning

We have studied the dependence of the machine learning performance on the number of training data set. The effects of data augmentation and cross validation are studied together. We use images from micromagnetic simulations to generate the training and the testing data sets. The number of the testing images is fixed to 1/10 of the training images. For a given training data set, four machines are developed (since the order of the learning process is randomized, the results can be different even though the training data set used is the same). The material parameters estimated from the four machines are used to obtain the mean value and standard deviation. The convolutional neural network (CNN) system is used, except noted otherwise, for machine learning.

The data size dependence of the machine learning performance (i.e. estimation error of the DM exchange constant) is shown in Supplementary Figure 2. Here we compare results with and without cross validation. For both cases, the estimation error saturates as the number of images used for training approaches 100,000 images. These results justify the data size used to develop the machine described in the main text.

Cross validation is carried out by splitting the images, generated from the simulations, into eleven groups with equal number of images (e.g. for 110,000 images, each group has 10,000 images). We use ten groups for training and the remaining one for testing to acquire the estimation error. This process is repeated 11 times: the group used for testing is changed each time. Since we develop 4 machines for a given training data set, the total number of machines for image recognition with and without cross validation is 44 and 4, respectively. The effect of cross validation is displayed in Supplementary Figure 2. The average estimation error with and without cross validation for 100,000 training images is ~0.045 mJ $m^{-2}$ and ~0.046 mJ $m^{-2}$, respectively. We thus find that cross validation does not significantly improve the estimation error.

The impact of data augmentation is shown by the colored symbols in Supplementary Figure 3. Here we use image rotation, shifting, and combination of rotation and shifting. The black solid circles show results without data augmentation. For each data point, we show average results from the four machines and cross validation (total machine number is 44).

Data augmentation with rotation is performed by taking a quarter of the images obtained from the simulations and create images that are rotated 90, 180 and 270 degrees. For example, among the 100,000 (10,000) training (testing) images output from the simulations, 25,000 (2,500) images are randomly selected and rotated 90, 180 and 270 degrees to generate a total of 100,000 (10,000) images. The estimation error of the DM exchange constant using such data augmentation (rotation) is shown by the red squares in Supplementary Figure 3.

The procedure for data augmentation with shifting is schematically sketched in Supplementary Figure 4. In micromagnetic simulations, we use a periodic boundary condition, which mimics the experimental condition (experimentally, ~2 μm × 2 μm size images are taken from samples that has a dimension of ~10 mm × 10 mm). The periodic boundary condition employed in the simulations allows data augmentation with shifting. The dimension of images used in the simulations is 2.048 μm × 2.048 μm. The shifting process is carried out by cutting the image into two pieces, shifting one half and displacing the other half in opposite directions. For vertical shifting (Supplementary Figure 4(a)), the bottom half of the image (2.048 μm × 1.024 μm) is shifted by 1.024 μm along the *y*-axis. The top half of the image (2.048 μm × 1.024 μm) is displaced by -1.024 μm along the *y*-axis. Similar procedure is performed for the horizontal shifting, which is displayed in Supplementary Figure 4(b). In addition, we augment data with a "vertical + horizontal" shifting, a combination of vertical and horizontal shifting: the process is sketched in Supplementary Figure 4(c). Using vertical, horizontal and vertical + horizontal shifting, three additional images are produced from the original image. Similar to data augmentation with rotation, we take a quarter of the images obtained from the simulations and create images that are shifted. The estimation error of the DM exchange constant is shown using the blue up triangles in Supplementary Figure 3.

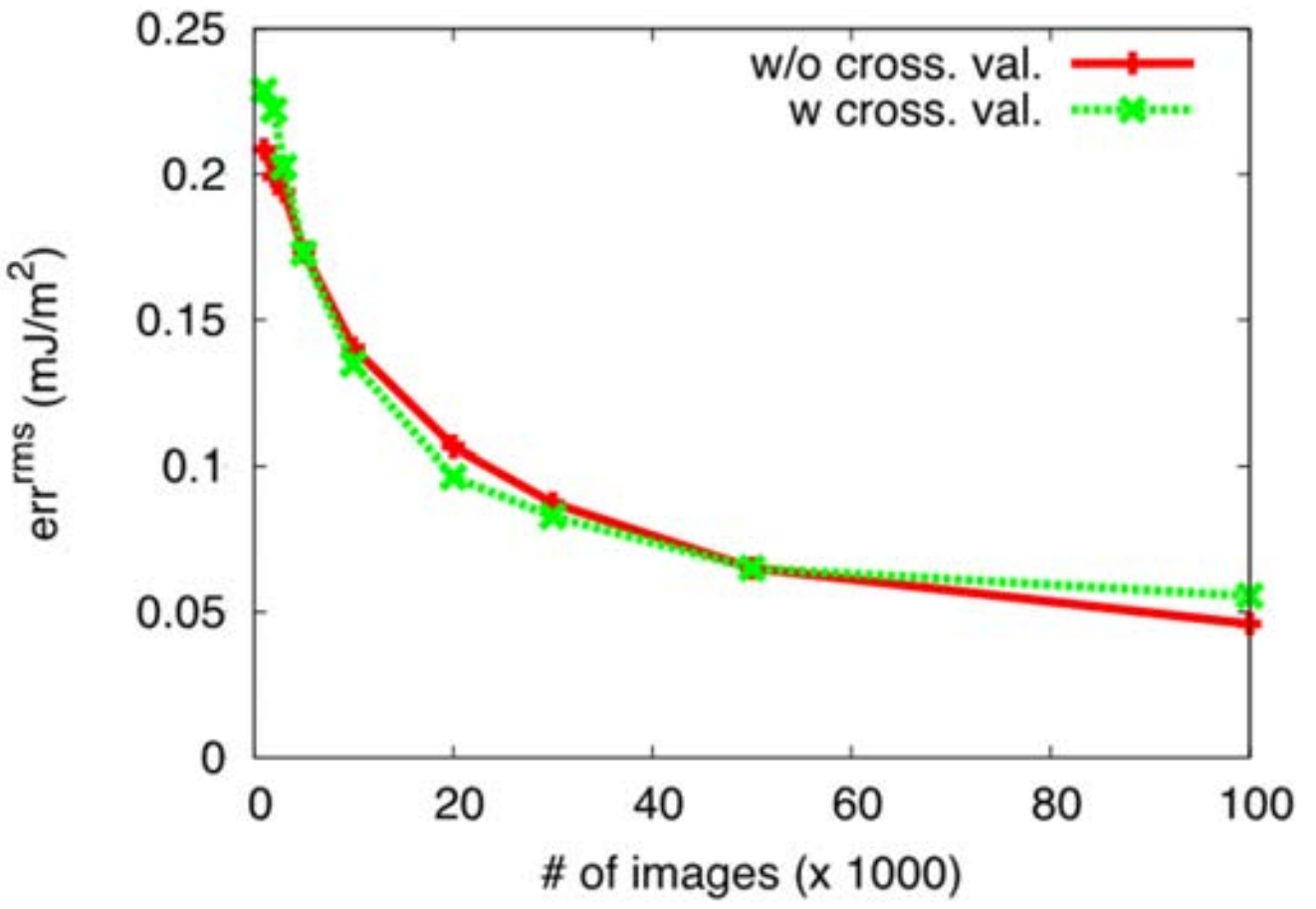

**Supplementary Figure 2: Assessment of the impact of cross validation.** Number of training images vs. the estimation error of the DM exchange constant ($D^{\rm est}$). The *x* axis represents the number of images used for training. Black squares: without cross validation, red circles: with 11-fold cross validation.

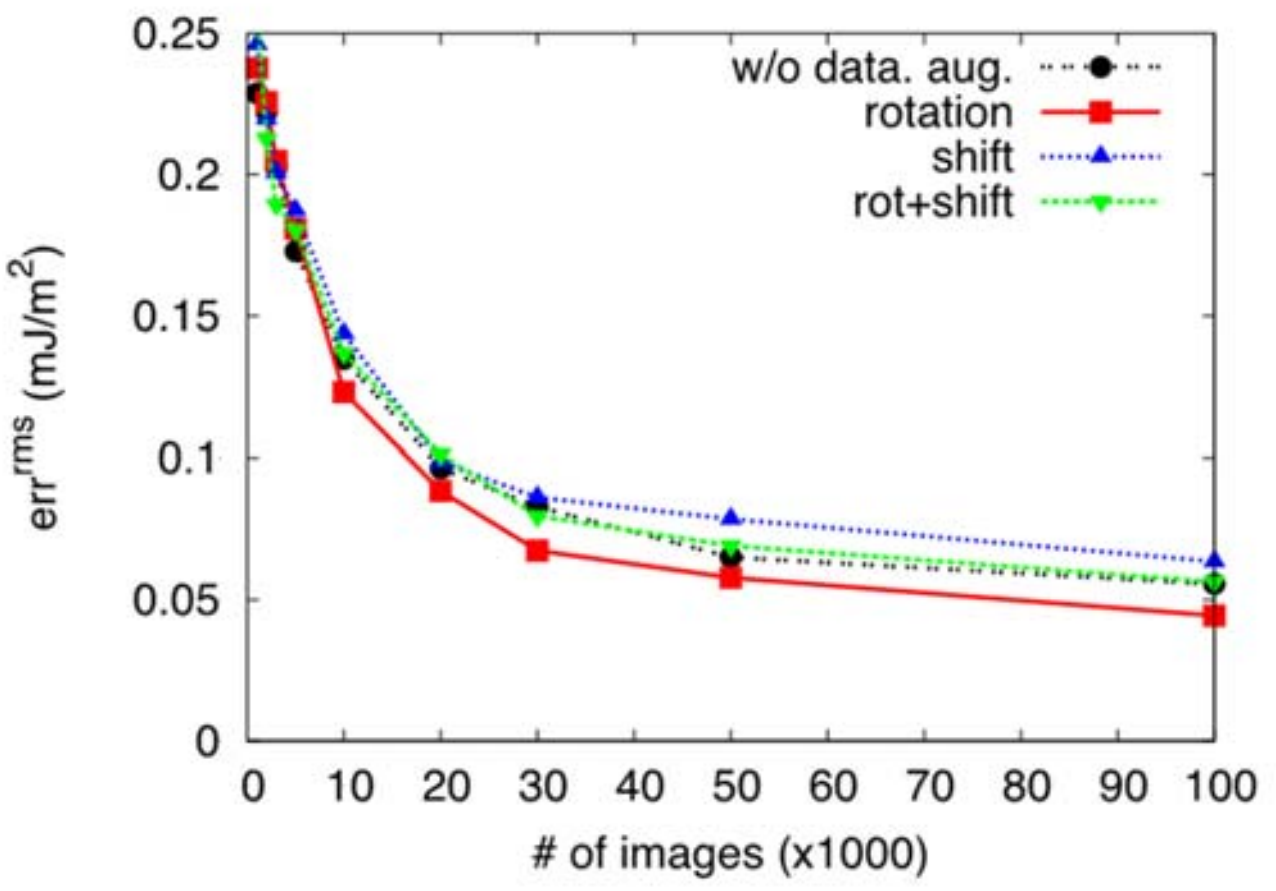

**Supplementary Figure 3: Evaluation of data augmentation.** The estimation error of the DM exchange constant ($D^{\rm est}$) plotted as a function of the number of images used for training. Black squares: without data augmentation, red circles: data augmentation with image rotation, blue up triangles: data augmentation with shifting, green down triangle: data augmentation with shifting + rotation. The 11-fold cross validation is implemented.

Finally, we can combine image rotation and shifting to generate images for data augmentation. From one image, we generate three additional images from rotation (90, 180, 270 deg) and another three additional images from shifting (vertical, horizontal, vertical + horizontal), which add up to total of 16 images (including the original image). The estimation error of the DM exchange constant using the augmented data of combined rotation and shifting is shown in Supplementary Figure 3, green down triangles.

The results shown in Supplementary Figure 3 suggest that data augmentation is effective in generating images for pattern recognition. For all three methods (rotation, shifting, rotation + shifting), the estimation error does not significantly increase from the data set without data augmentation. Using data augmentation with rotation + shifting, one may reduce the training/testing data size by a factor of 16.

In addition to the CNN system describe in the main text, a simple residual network (RN) system (https://dl.sony.com/project/, search by technique, ResNet, tutorial.basics.12_residual_learning) has been tested for comparison. In Supplementary Figure 3, the average estimation error for 100,000 training images without cross validation was ~0.045 mJ $m^{-2}$. Under the same condition, the RN system provides an estimation error of ~0.032 mJ $m^{-2}$. As an initial screening, the RN system seems to perform better, however, with the cost of computation time, which is nearly two times larger using our computing resources.

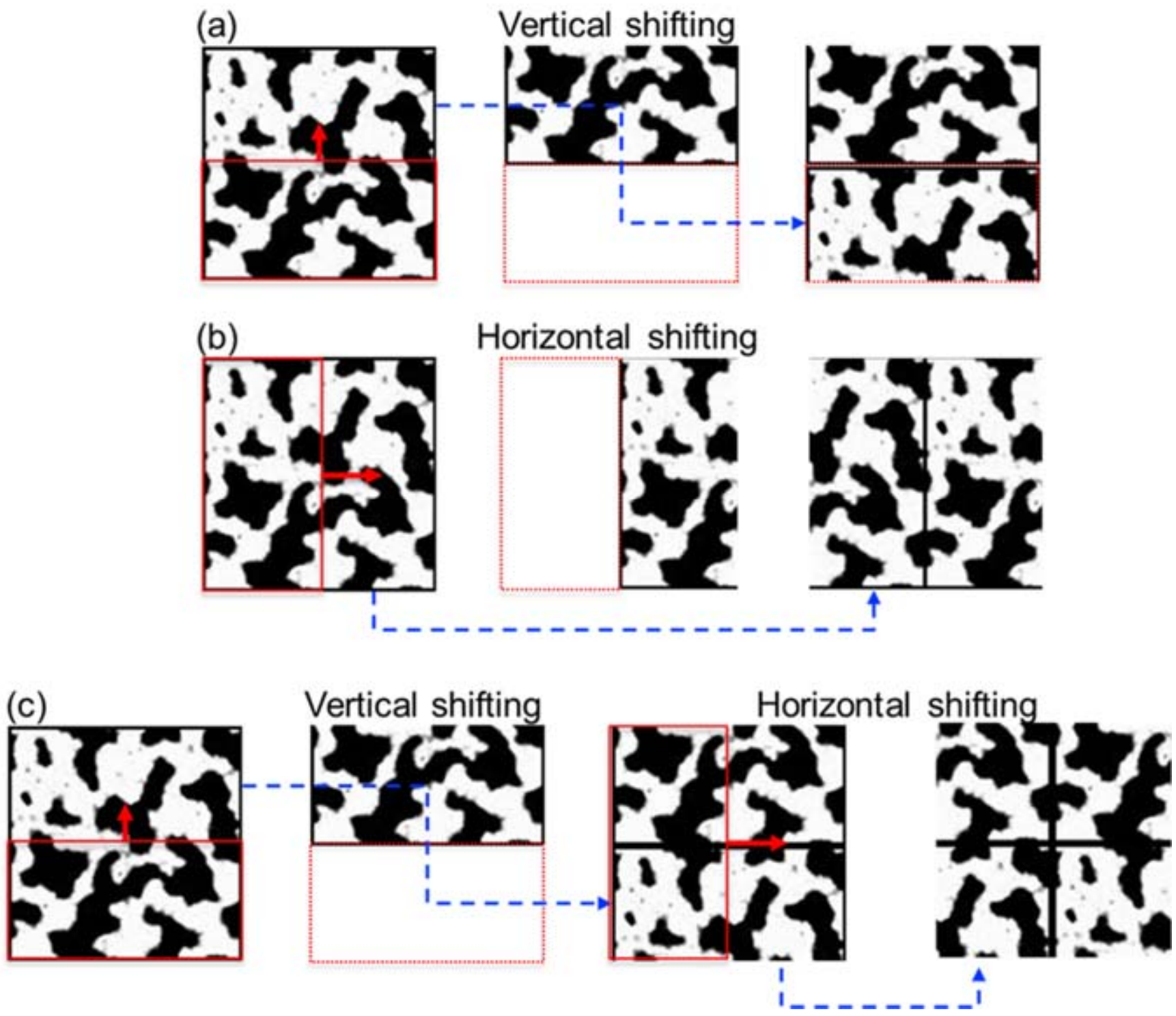

**Supplementary Figure 4: Schematic illustration of the image shifting procedure.** (a-c) Vertical (a), horizontal (b) and vertical + horizontal (c) image shifting processes are illustrated. A periodic boundary condition applied to the images in the simulations, allowing image shifting without causing discontinuity from the edges.

**Supplementary References**